\documentstyle[12pt,epsfig]{article}
\date{}
\begin{document}
\title{Vacuum polarization energy losses of
high energy cosmic rays.} 
\author{V.A.Maisheev \thanks{E-mail maisheev@mx.ihep.su} \\
{\it Institute for High Energy Physics, 142284, Protvino, Russia }}

\maketitle
\def\arcctg{\mathop{\rm arccot}\nolimits} 
\def\ch{\mathop{\rm ch}\nolimits}
\def\sh{\mathop{\rm sh}\nolimits} 
\def\Im{\mathop{\rm Im}\nolimits}
\def\Re{\mathop{\rm Re}\nolimits}
\def\sign{\mathop{\rm sign}\nolimits}
\def\div{\mathop{\rm div}\nolimits}
\def\rot{\mathop{\rm rot}\nolimits}
\def\grad{\mathop{\rm grad}\nolimits}
\begin{abstract}
 The process of the vacuum polarization energy losses of 
high energy cosmic rays propagating in the extragalactic space 
is considered. The process is  due to the polarization of Cosmic
Background Radiation by a moving charged particle. With the goal of
the description of the process, the photon mass, refractive indices
and permittivity function for low and high energy photons are found.
Calculations show the rather noticeable level of the energy losses
for propagating protons with the energies more than $10^6 - 10^7 GeV$.
The influence of the polarization energy losses on  propagation of
cosmic rays is discussed.
\end{abstract}

\section{ Introduction}
 The problems of the origin and propagation of  high energy cosmic rays
is widely discussed in recent years (see, for example \cite{NW,BGG,AP} and
literature therein). The photoproduction of mesons and $e^\pm$-pairs
in the Cosmic Background Radiation
was considered as the main reason of the energy losses of charged particles
propagating in the extragalactic space\cite{BGG,Lee,St}. 
However, the simulations 
of the spectral distributions of cosmic rays on this basis do not provide 
a close  agreement with observed data \cite{BGG}, 
and some peculiarities in the spectrum 
like the "knee" \cite{K} and  GZK-cutoff \cite{KG,ZK} 
do not find recognized explanation. 
In this connection the
determination  of  another reason of the energy losses is important and
interesting.

 In this paper we consider the possible mechanism by which  cosmic rays
lose their energy in the extragalactic space. It is
the polarization energy losses  of a charged particle moving in 
the electromagnetic vacuum. In the presence of an 
external  electromagnetic field the polarization of vacuum was considered
first in the pioneer papers \cite{JS}. 
The vacuum polarization  leads  to different effects \cite{LL1} 
such as the nonlinearity
of the Maxwell equations, appearance of nonzero photon mass, birefringence  
of light, etc. The description of the various methods in the QED of
vacuum one can find in literature ( see \cite{FGS} and references therein).
In parallel with other methods  the traditional method, 
based on the introduction
of the permittivity tensor, may be used  for the effective description  of 
different phenomena in the electromagnetic vacuum. In this case the vacuum
described by the Maxwell equations which are similar to equations
of the classical electrodynamics for continuous media\cite{LL,AG}. 
As an example, one can point
out the well known  low energy permittivity and permeability tensors for
 constant electric and magnetic fields \cite{LL1}. In paper \cite{BK} 
the connection 
between the polarization  and permittivity tensors is established.
The using of the permittivity tensor  is
convenient because this approach allows to consider the QED vacuum and 
matter with the unit point of view.

\section{ The mass of photon, propagating in an isotropic photon gas}     
 The photon mass $m_\gamma$ in an isotropic medium is defined by 
 the refractive index $\tilde{n}$  of monochromatic photons 
propagating in this medium  
\begin{equation}
m_\gamma^2 c^4 =2 E_\gamma^2 (1- \tilde{n}(E_\gamma))
\label{1}
\end{equation}
where $E_\gamma$ is the photon energy.
From this equation
one can see the equivalence of  finding of the photon mass and 
the refractive index. 
Note that in the general case the refractive index is a complex value. 
However, in a transparent medium  its  imaginary part is a negligibly small
quantity.

 Our calculations of photon mass in an isotropic photon gas are based on 
the results of papers
\cite{MV2,MV3}, where propagation of $\gamma$-quanta in the field of 
monochromatic and dichromatic laser waves was considered. 
In these papers the permittivity
tensor was derived for such media. Knowing this tensor one can find the
refractive indices of photon and other characteristics of the propagation 
process. The imaginary components
of the tensor are connected with the energy losses due 
to $e^\pm$-pair production
in the laser wave. The real components of the permittivity
tensor are derived with the help of dispersion relations.    
In the case of an isotropic photon gas the energy losses one can 
describe by the following
relation (see, for example\cite{PM}):
\begin{equation}
L_{\gamma \gamma}^{-1}(E_\gamma) = {1 \over {8E_\gamma^2}} 
\int_{\epsilon_{min}}^{\infty}{d\epsilon {n(\epsilon) \over \epsilon^2}
\int_{s_{min}}^{s_{max}}{\sigma_{\gamma \gamma} sds}}
\label{2}
\end{equation} 
where $\L_{\gamma\gamma}$ is the reciprocal of the mean free path for
collisions, $n(\epsilon)$ is the differential photon gas number density
for the photon energy  equal to $\epsilon$,
$s$ is the center of momentum frame energy squared,
$\sigma_{\gamma \gamma}(s)$ is the total cross section of $e^{\pm}$-pair
production,
$E_\gamma$ is the energy of photon propagating in a medium,  
$s_{min}=4 m^2c^4$, $s_{max}= 4\epsilon E_\gamma$, 
$\epsilon_{min}=s_{min}/4E_\gamma$, m is the electron mass and 
$c$ is the speed of light in the vacuum.
All relations in this paper will be valid for the laboratory coordinate
system, which determined as the system where the mean momentum of
the background photons is equal to zero.    
From Eq.({2}) one can find the imaginary part of refractive index $\tilde{n}$ 
\begin{equation}
Im(\tilde{n}(E_\gamma)) 
= {{L_{\gamma\gamma}^{-1} \hbar c } \over {2E_\gamma}}
\label{3}
\end{equation}
The explicit form of this relation  is:
\begin{equation}
Im(\tilde{n}(E_\gamma))= {{\pi r_e^2 \lambda_e}\over{mc^2}} 
\int_{\epsilon_{min}}^{\infty}{ {\cal{F}}_{im}(z_m) n(\epsilon) \epsilon   
d\epsilon}, 
\label{4}
\end{equation}
\begin{equation}
{\cal{F}}_{im}(z_m) = z_m^3 \int_{z_m}^{1}{[(1+z-z^2/2 )
\ln{{1+\sqrt{1-z}}\over{1-\sqrt{1-z}}}- \sqrt{1-z}(1+z)}]{dz\over z^2}  
\label{5}
\end{equation}
where $r_e$ and $\lambda_e$  are the classical radius of electron and its 
Compton wavelength, $z=4m^2c^4/s$ and
$z_m=4m^2c^4/s_{max}= m^2c^4/(\epsilon E_\gamma)$.
The corresponding real part of the photon refractive index is:
\begin{eqnarray}
Re(\tilde{n}(E_\gamma))-1 ={{2r_e^2 \lambda_e}\over {mc^2}} \int_0^{\infty}
{{\cal{F}}_{re}(z_m)n(\epsilon)\epsilon d\epsilon}, 
\label{6}
\end{eqnarray}
where function 
${\cal{F}}_{re}(z)$ is determined by the following relation:
\begin{equation}
{\cal{F}}_{re}(z_m)= z_m^3\int_{z_m}^\infty \Phi(z){dz \over z^2}
\label{7}
\end{equation}
\begin{equation}
\Phi(z)=
\cases{ 
-3 -(1+z-{z^2\over 2}){1\over 4} L^2_- 
-(1-(z+{z^2\over 2})){1\over 4}L^2_+ + \cr
+{{(1+ z )\sqrt{1-z}}\over{2}}L_- -{{( z-1)\sqrt{z+1}}\over {2}}L_+ 
+ {\pi^2 \over 4}(1+ z-{{z^2}\over 2}),\; 0<z \le 1, \cr
-3  +(1+ z-{z^2\over 2}) \arcctg^2(\sqrt{z-1}) 
-(1- (z+{z^2 \over 2})){1\over 4}
L^2_+ + \cr
+ (1+ z)\sqrt{z-1} \arcctg\sqrt{z-1} -{{( z-1)\sqrt{1+z}}\over{2}} L_+ 
,\; z>1.  \cr }
\label{8}
\end{equation}
The functions $L_{-}, \, L_+$ are equal to:
\begin{equation}
L_-=\ln{{1+\sqrt{1-z}}\over{1-\sqrt{1-z}}}, \quad 
L_+=\ln{{\sqrt{1+z}+1}\over{\sqrt{1+z}-1}}. \nonumber  
\label{9}
\end{equation}
The obtained here relations for refractive indices are valid for low
and high energy photons propagating in a photon gas. 
Now one can calculate the refractive indices of the photon propagating
in the extragalactic space. As the first approximation of this medium
one can take the model of the space which is filled by the Cosmic 
Background Radiation. In this case the number density is defined
by well known  equation:
\begin{equation}
n(\epsilon) d\epsilon ={1 \over { \pi^2 c^3 \hbar^3}}
{{\epsilon^2 d\epsilon}\over
{\exp(\epsilon/T)-1}}\,,
\label{10}
\end{equation}
where $T$ is the temperature.
For low energy photons one can find the following
relation for real part of the refractive index:
\begin{equation}
Re(\tilde{n}(0))-1 ={{44 r_e^2 \lambda_e}\over {135 mc^2}} \int_0^{\infty} 
{n(\epsilon)\epsilon d\epsilon}  =
{{44\pi^2 \alpha^2} \over 2025} {{T^4}\over {m^4c^8}}\,.  
\label{11}
\end{equation}
The real and imaginary parts of refractive indices are connected by the
dispersion relation:
\begin{equation}
 Re(\tilde{n}(E_\gamma)) - 1 = {2 \over \pi} {\cal{P}}\int_0^{\infty}
 { {\hat{E}_\gamma Im(\tilde{n}(\hat{E_\gamma})) d \hat{E}_\gamma} \over
 {\hat{E}_\gamma^2 - E_\gamma^2}}
\end{equation}
At present temperature $T=2.726^o \, K$ the real part of refractive index
 for low energy photons
is $5.1 \, 10^{-43}$. These results are in agrement
with the similar calculations \cite{Bart,LPT,MT} of the low energy refractive
indices in an isotropic photon gas.

Fig.1 illustrates  the results of calculation of the photon 
refractive indices in the simplest model of the extragalactic medium
(in the Cosmic Background Radiation). One can see that the real part 
is nearly constant in the wide energy range from 0 till $2\, 10^5$  GeV
with the flat maximum ($\approx  6\, 10^{-43}$)  
at  $E_\gamma\approx 3\, 10^5 GeV$.
At energies more than $2.5 \, 10^6 GeV$ the real part of index is negative. 
The imaginary part has maximum value 
$ \approx 4\, 10^{-43}$ at $E_\gamma= 8\, 10^5 GeV$. 

Knowing the refractive indices one can find the energy dependence of
the permittivity function of a photon gas. However,  this function 
is not uniquely defined and it depends on the relation between magnetic 
induction vector $B$ and intensity of magnetic field $H$ (see \cite{LL,AG}).
In  \cite{MV2} the Maxwell equations are written with $H = B$ and in this 
case the permittivity function $\varepsilon$ is   
\begin{equation}
\varepsilon(E_\gamma)=
\tilde{n}^2 \approx 1 +2 Re(\tilde n -1) +2iIm(\tilde{n}). 
\label{13}
\end{equation}
\section{Energy losses of particle moving in a medium}
Propagating in the Cosmic Background Radiation charged cosmic rays 
produce polarization of this medium.
As result the cosmic rays lose the initial energy. This process 
is similar in many respects to the ionization losses of energy of charged
particles moving in  matter. It is clear that the vacuum polarization   
is  many times more weak process, however it may be noticeable  
on a large distances of particle propagation in such a specific medium as
the extragalactic space.
 
Our consideration  will be based on the  Landau theory 
 of ionization energy losses of relativistic particles in a matter \cite{LL}.
However, a small adaptation of the theory will be done with the goal of the 
elimination of divergences in the final result.

The Maxwell equations in a medium  have the following form:
\begin{eqnarray}
\div{{\bf{B}}}= 0, \qquad \rot{{\bf{E}}} 
= -{1\over c} {\partial{{\bf{B}}} \over \partial{t}}, \label{14} \\
\div{\hat{\varepsilon}{\bf{E}}} = 4\pi \rho ,\,\,\, 
\rot{{\bf{B}}}=  {1\over c} 
{\partial{\hat{\varepsilon}}{\bf{E}} \over \partial{t}}+
{4\pi \over c}{\bf{j}}.  
\label{15}
\end{eqnarray}  
where ${\bf{E}}$ is the intensity of electric field and $\bf{B}$ is
the  magnetic induction vector, $\rho$ and ${\bf{j}}$
are the charge and current densities, $\hat{\varepsilon}$ 
is the permittivity operator (see\cite{LL}),   t is the time.
  
  The charge and current  distributions one can take in the following form:
\begin{equation}
\rho =e\rho({\bf{r}} - {\bf{v}}t), \quad 
{\bf{j}}= e\rho({\bf{r}}- {\bf{v}}t){\bf{v}},
\qquad \int_V{\rho(\bf{r})d{\bf{r}}}=1, 
\label{16}
\end{equation}
where $e$ is the particle charge.
The field potentials we write in the usual form:
\begin{equation}
 {\bf{B}}= \rot{{\bf{A}}},\qquad 
{\bf{E}}= -{1 \over c} {\partial{{\bf{A}}} \over \partial{t}}
- \grad \varphi 
\label{17}
\end{equation}
In accordance with \cite{LL}  we use the further condition for potentials:
\begin{equation}
\div{{\bf{A}}}+{1\over c} 
{\partial{\hat{\varepsilon}}\varphi \over \partial{t}} =0.
\label{18}
\end{equation}
The substitution Eq({17}) in Eq({15}) allows to get 
the following relations
for potentials:
\begin{eqnarray}
\Delta{\bf{A}} -{\hat{\varepsilon}\over c^2} 
{\partial^2{\bf{A}} \over \partial{t^2}}
= -{4\pi \over c} e\rho({\bf{r}} - {\bf{v}}t) {\bf{v}}, \label{19} \\ 
\hat{\varepsilon}(\Delta\varphi - {\hat{\varepsilon}\over c^2}
{\partial^2\varphi \over \partial{t^2}})= - 4\pi e \rho({\bf{r}}-{\bf{v}}t).   
\label{20}
\end{eqnarray} 
These equations for Fourier components have the form:
\begin{eqnarray}
k^2 {\bf{A_k}}+ {\hat{\varepsilon}\over c^2} 
{\partial^2{{\bf{A_k}}} \over \partial{t^2}}=
{4\pi \over c} eF_{\bf{k}} {\bf{v}} \exp({-i{\bf{kv}}t}), \label{21} \\
\hat{\varepsilon}(k^2 \varphi_{\bf{k}} + {\hat{\varepsilon} \over c^2}
{\partial^2\varphi_{\bf{k}} \over \partial{t^2}})=
4\pi e F_{\bf{k}} \exp{(-i{\bf{kv}}t)}.     
\label{22}
\end{eqnarray}
where $F_{\bf{k}}=\int{\rho({\bf{r}})\exp({-i{\bf{kr}}})d{\bf{r}}}$.
One can see that the functions ${\bf{A_k}}$ and 
$\varphi$ depend on the time as
$\exp{(-i{\bf{kv}})}$. As it shown in \cite{LL},  
the result of action of $\hat{\varepsilon}$-operator 
on $\exp{(-i\omega t)}$-
function  is its multiplication on $\varepsilon(\omega)$-function.
Now from Eqs.({21}-{22}) one can get:
\begin{eqnarray}
{\bf{A_k}}= {4\pi e \over c} \,
{F_{\bf{k}} {\bf{v}} \over k^2 -{(\bf{kv}})^2 \varepsilon({{\bf{kv}}})/c^2}
\exp{(-i{\bf{kv}}t)}, \label{23} \\
\varphi_{\bf{k}} = {4\pi e \over \varepsilon({{\bf{kv}})}}\,
{F_{\bf{k}} \over k^2-{(\bf{kv}})^2 \varepsilon({{\bf{kv}}})/c^2}
\exp{(-i{\bf{kv}}t)}.
\label{24}
\end{eqnarray} 
The Fourier component of the intensity of electric field has the
following form:
\begin{equation}
{\bf{E_k}}={i{\bf{kv}} \over c} {\bf{A_k}} - i{\bf{k}}\varphi_{\bf{k}}.
\label{25}
\end{equation}
Now we find the damping force, which act on the extended charge. 
For this purpose we use the following well known relation:
\begin{equation}
{d\over dt} ({{\bf{D E}}+B^2 \over 8\pi}) = -{\bf{jE}} - \div{\bf{S}}, \quad 
{\bf{S}}= {c \over 4\pi} [{\bf{EB}}],
\label{26}
\end{equation}
where $\bf{D}$ is the electrical induction vector.
 After integration over some volume $V$ one can get:
\begin{equation} 
{d{\cal{H}} \over dt}= - e{\bf{v}} 
\int_V{\rho({\bf{r}}){\bf{E}}({\bf{r}})d{\bf{r}}}     
-\oint_\sigma{S_nd\sigma}, \qquad
{\cal{H}}=\int_V{{{{ {\bf{DE}}+B^2}\over 8\pi}\,dV}},
\label{27}
\end{equation}
where $\sigma$ is the bounding surface. 
It is clear, that at ${\bf{r}}\rightarrow \infty$
the surface integral  tends to zero. 
Then the damping force has the following  
form:
\begin{equation}
{\bf{  {\cal{F}}  }} = - {e{\bf{v}}\over v} 
\int_V{\rho({\bf{r}}){\bf{E}}({\bf{r}})\,d{\bf{r}}}
\label{28}
\end{equation}
Obviously  that  Fouirier component of electric intensity 
has the form: ${\bf{E_k}}= {\bf{\Phi}}({\bf{k}}) \exp{-i{\bf{kv}}t}$,
where ${\bf{\Phi}}({\bf{k}})$ is the known function (see Eqs.({23}-{25})).
Then we can write the following relation:
\begin{eqnarray}
 e {\bf{v}}\int_V{\rho({\bf{r}}-{\bf{v}}t){\bf{E}}\, d{\bf{r}}} = 
\qquad \qquad \qquad  \qquad \qquad \qquad \quad \\ 
{e{\bf{v}} \over (2\pi)^3} 
\int{
\{\int_V\rho({\bf{r}}-{\bf{v}}t)\exp{i{\bf{k}}
({\bf{r}} - {\bf{v}}t)} d{\bf{r}}\} 
 \Phi({\bf{k}})d{\bf{k}}
}=
{e{\bf{v}} \over (2\pi)^3}  \int{F_{\bf{k}} {\bf{\Phi}}({\bf{k}})\, 
d{\bf{k}}}, 
\nonumber
\label{29}
\end{eqnarray}
Obviously, the damping force is directed against the velocity of particle.
Let us name $k_xv = \omega$ and $ q= \sqrt{k_y^2+k_z^2}$, where
 $x$-coordinate axis
is along the direction of motion, $\omega$ is the photon frequency.  
Now one can obtain the following equation for the absolute value of 
the damping force:
\begin{equation}
{\cal{F}}={ie^2\over \pi} \int_{-\infty}^{\infty} {\int_0^\infty
{{F^2_{\bf{k}}(k) ({1\over v^2}-{\varepsilon \over c^2})\omega q \,dqd\omega}
\over
{\varepsilon[ q^2 +\omega^2({1\over v^2}-{\varepsilon \over c^2})]}
}}
\label{30}
\end{equation}
One can see that these relation is differ from similar relation in \cite{LL}
by $F^2(k)$-multiplier. 
 This multiplier is the charge distribution of the 
moving  particle in ${\bf{k}}$-space  
(the axial symmetry of the charge distribution around x-axis is assumed).
For next we can write that 
$\varepsilon \approx  1 +\Delta\varepsilon' +i\Delta\varepsilon'',\,
|\Delta\varepsilon'| <<1, \Delta\varepsilon'' << 1$. Then we 
extract the real and imaginary parts of the integrand function 
in Eq.({30}). Thus,
we get the sum of two integrals: $I_1 + iI_2$. Besides, it needs to take
into account that $\Delta\varepsilon'$ and $\Delta\varepsilon''$
are even and odd functions of the frequency. 
One can show that $I_2=0$, and therefore the damping force is:
\begin{equation}
{\cal{F}}= {2e^2 \over \pi v^2} \int_0^\infty{\int_0^\infty{
{F^2(q^2 + {\omega^2\over v^2\gamma^2}) \Delta{\varepsilon''}(\omega) 
(q^2 + {\omega^2 \over \gamma^4 v^2}) \omega q \, d\omega dq 
\over
(q^2 +{\omega^2 \over v^2 \gamma^2}- 
{\omega^2 \Delta\varepsilon' \over c^2})^2 +
{\omega^4 \Delta\varepsilon''^2 \over c^4}} 
}}
\label{31}
\end{equation}   
where $\gamma$ is the Lorentz factor of the particle. Besides, 
in this equation
$F$ is the charge form factor of the particles (i.e. the charge distribution
in ${\bf{k}}$-space in the particle  rest frame).
 The Eq.({31}) for the damping force
is final. The two integrals in this equation are converged. 
The integral over
$q$ is converged due to the form factor of the particle, 
and the integral over
$\omega$ is converged due to relation:$ \varepsilon \rightarrow 1$
at $\omega \rightarrow \infty$. 
\section{Energy losses of cosmic rays}   
For cosmic rays Eq.({31}) one can simplify. It is well known that the maximum 
observed energy of cosmic rays is less than $10^{12} \, GeV$. 
One can see that energy losses depend on the real and imaginary parts of the
permittivity function.  However, one can neglect of this dependence in   the
denominator. It is really that $\gamma^{-2} >>|\Delta\varepsilon'|,\,
\Delta\varepsilon''$(see Fig.1 and \cite{Dr}). Now one can write 
the damping force for cosmic rays in the following form:
\begin{equation}
{\cal{F}}=
{2e^2 \over \pi \hbar^2v^2}\int_0^\infty{\int_0^\infty{
{F^2(\bar{q}^2+ {E_\gamma^2 \over v^2\gamma^2}) \varepsilon''(E_\gamma)
(\bar{q}^2 + {E\gamma^2 \over v^2 \gamma^4}) 
E_\gamma \bar{q} \, dE_\gamma d\bar{q} 
\over
(\bar{q}^2 + {E_\gamma^2 \over \gamma^2 v^2})^2}
}}\,
\label{32}
\end{equation}
where $\bar{q} =\hbar q$ and $E_\gamma = \hbar \omega$. Here we use   
relation: $\varepsilon'' = \Delta\varepsilon''$. 
For the next calculations
we take the empirical electric proton form factor \cite{P}:
\begin{equation}
F(Q)= (1+{Q^2 \over M^2_V})^{-2}.
\label{33}
\end{equation}
where $Q$ is the three-dimensional transfer momentum,
 the empirical constant $M^2_V = (0.84 GeV/c)^2$. Then we find the
integral over $\bar{q}$ and get the following relation:
\begin{equation}
{\cal{F}}= {e^2 \over \pi \hbar^2 v^2} \int_0^\infty{  
\varepsilon''(E_\gamma)(S_1(E_\gamma)+S_2(E_\gamma))E_\gamma\, dE_\gamma 
}\,
\label{34}
\end{equation}
\begin{eqnarray}
S_1(E_\gamma)= -\ln(xy) -{y^3\over 3} - {y^2\over 2} - y, \label{35} \\
S_2(E_\gamma)= 
x({1\over \gamma^2}-1) ( 4\ln(xy) + {y^3 \over 3} + y^2 +3y +{1\over x}), 
\label{36}\\
x= {E_\gamma^2 \over \gamma^2 v^2 M^2_V}, \qquad \qquad \qquad \qquad
y={1 \over 1+x}.
\label{37}
\end{eqnarray}
It is helpful to obtain the inexact but simple estimation relation for
the damping force. It is possible to make for the large $\gamma$-factors
 $(> 10^8)$  of the cosmic rays. This relation is:
\begin{equation}
{\cal{F}}\sim { e^2 \over \pi \hbar^2 v^2}
\ln({\gamma M_V v\over <E_\gamma>})
\int_0^{E_{\gamma,b}}{\varepsilon''(E_\gamma)E_\gamma dE_\gamma}
\label{38}
\end{equation}
where  $<E_\gamma>$ is some value of the $\gamma$-quantum energy range, 
where the integrand has noticeable quantities, $E_{\gamma,b}$ is the boundary
value of the energy range and it is satisfied the condition 
$E_{\gamma,b} \gg  \gamma M_V v$. For example,
one can define the $\bar{E_\gamma}$ by the following equation:
\begin{equation}
\bar{E_\gamma}=\int_0^{E_{\gamma,b}}{\varepsilon''(E_\gamma) 
E_\gamma dE_\gamma}\,\, /
\,\,\int_0^\infty{\varepsilon''(E_\gamma) dE_\gamma}.
\label{39}
\end{equation}
According to calculations $<E_\gamma>\approx 8\,10^7 GeV$ and the quantity
of the integral in the numerator of Eq.({39}) is equal to 
$\approx 10^{-28} GeV^2$ at $E_{\gamma,b} \sim 10^{13} \,GeV$.
The choice of $E_{\gamma,b}$ in the energy range from 
$10^{10}$ till $10^{14} \, GeV$ is weakly changed the result.
 Then we get the
following estimation:
\begin{equation}
{\cal{F}} \sim { \alpha Z^2 \over \pi \hbar c} 10^{-28}
\ln{\gamma M_V c \over <E_\gamma>},  
\label{40}
\end{equation}        
where $Z$ is the particle charge in the elementary one units.
From here one can see that energy losses of high energy cosmic rays 
(protons)
is about some hundreds of GeV per light year. This value is the same
order as the $e^\pm$-pair production in the Cosmic Background Radiation.
Fig.2  illustrates the calculation of vacuum polarization energy losses
of protons in accordance with Eq.({34}).

\section{Discussion}
The results of calculations of the vacuum polarization energy losses of
the high energy cosmic rays show that losses are reasonably large, and,
because of this, it is necessary to take into account this process for
true description of the  particle propagation in the extragalactic space. 
 Fig.3 illustrates the relative polarization energy losses of protons
 and iron nuclei in the extragalactic space. The results of calculation
 \cite{BGG} of the energy losses  due to the photoproduction processes
 ($p\,+\,\gamma \rightarrow \, p\,+\,e^+\,+\,e^-,\, p\,+\,\gamma \rightarrow 
  \, p\,+\, \pi^0,...$) are  also shown on the figure.
 One can see the following peculiarities of the considered process:

  i) the spectral distribution of polarization energy losses are rather
 broad, and the flat maximum of the relative losses  is  at 
 $E_{p,max} \approx 4\,10^7 GeV$ and $E_{Fe,max} \approx 1.4\, 
 10^{10} GeV$ for
 protons and iron nuclei, correspondingly; 

  ii) the relative value of the polarization energy losses has the same
 order as the losses due to the photoproduction processes (at proton
 energies in the range $ 10^{7}-10^{10} \, GeV$);
 
 iii) the strong drop of polarization losses take place for particle energies
  $\ll E_{p,max},\, E_{Fe,max}$; 

  iv) the spectral behavior of proton and iron polarization losses  
 is the same, but there is valuable energy shift between the curves
 on Figs.2-3. As result at energies $< \,10^9\, GeV$ the polarization 
 losses of protons is more, than ones for iron nuclei. 
 The reverse situation is observed at energies $>\,  10^9 \, GeV$;

 v) our calculations show that the points of inflection are exist for the
 both curves on Fig.2 (when $d^2{\cal{F}}/d^2E =0$). It takes  place
 when the energies of particles are equal to $1.5\,10^7 $ 
 and $1.4 \, 10^{10}\, GeV$
 for protons and iron nuclei, correspondingly.
 
 The behavior of energy losses for iron nuclei 
one can understand if taken into account that the losses are proportional
to the square of charge but the form factor cuts of the value of 
damping force more strong for iron, than for proton (the constant 
$M_V = 0.13 GeV/c$ ).        

 Now we make attempt to explain some experimental data in the observation
of the cosmic ray spectrum. The absence of the clear  
photoabsorbtion threshold
in the spectrum \cite{BGG, St} one can explain by the common continuous
character of the summary energy losses. Really, from fig.3 we see that
the polarization energy losses sewed together with the photoabsorbtion ones.
It means that clear photoproduction threshold can not be detected on this
background.  

 The "knee"-effect is another misunderstand area in the spectrum of the
cosmic rays. One can suggest that the "knee" is a point where the
character of cosmic ray propagation is changed. At energies above the
"knee"-level the particles lose their relative energy and after passing
the "knee"-point  the energy losses are decreased  and accumulation particles
in this area takes  place.
 
Some remarks concerning the composition of cosmic rays. The composition
is dominated  by protons at the lowest energies, and then the fraction of
light nuclei increases with energy\cite{K,Bu,Be}. 
However, at energies $> 5\, 10^9 GeV$
the protons is again dominated. We can see on Fig.2 that at low energies
the polarization losses of protons in many times exceed the same losses for
iron nuclei. On the other hand, at energies  $ >5\,10^9$ polarization
losses for  iron nuclei are large. In particular, the composition 
of cosmic rays
is determined by lifespan  in which a cosmic particle keeps the energy.  
Note, the "knee" is observed only for proton fraction, and it is absent
for iron nuclei with energies $\le 10^9 GeV$. From our point of view 
this fact is obvious. We believe that the iron "knee" is at energies more 
than $5\, 10^{9} \, GeV$ near the point of inflection for 
the iron energy losses curve.
       
In the case if the considered here mechanism of vacuum polarization energy
losses is true the following  statement take place: the initial number
(or the speed of production)
of cosmic rays ( with energies above the "knee"-point) is more, than it is
commonly supported.   
  
These our conclusions have the qualitative character. However, no doubt
they may be tested by simulation of the propagation process  of cosmic 
rays. 

In conclusion, we have touched on the nature of the considered phenomenon.    
The using of methods of the electrodynamics of the continuous media
allow  to describe mathematically the mechanism of the vacuum 
polarization energy losses of relativistic particles without detailed
description of the  primary processes, which are responsible for this
phenomenon. From this standpoint the moving charged particles polarize the
medium. It requires some energy and the particles give back its to
the medium. In the case of the electromagnetic vacuum this energy go
into creation of virtual $e^\pm$-pairs. According to Eq.({39}) 
the spectrum
of the virtual pairs is the rather broad and its upper bound is near
the energy of particles. Although the quantities of the permittivity
function $\varepsilon''$ are rather small, the noticeable value
of the energy losses is generated due to broad and high energy spectrum
of virtual pairs in the laboratory coordinate system. 
It is obvious that these pairs are low energetic
in the rest frame of the cosmic charged particle.  Then the real photons
from the Cosmic Background Radiation and virtual ones can 
interact effectively
in between at the condition $ \epsilon E_{\gamma v}> m^2 c^4$,
where $E_{\gamma v}$ is the energy of the virtual state. 

It should be noted, that we do not consider the influence of the red shift
on the energy losses. It needs to make in the case when propagation 
of the cosmic rays is investigated on  long distances. 
Besides, we think that the contribution in the vacuum polarization
energy losses of cosmic rays from another background fields is
possible. 
\section{Conclusion}
On the  basic of determination  such characteristics as the photon mass,
refractive indices and permittivity function in the Cosmic Background
 Radiation the vacuum polarization losses of high energy cosmic rays
are considered. The calculations show the high level of these 
losses for protons with energies more than $\approx 10^7 GeV$.
The proposed mechanism of losses leads to a revision of the existing 
propagation models of cosmic rays. With our point of view 
the propagation of the high energy cosmic rays in the extragalactic
space is the dynamic process to a greater extent than it is expected. 
Experimental and theoretical ivestigations 
of these processes will help to understand the nature and origin of 
the cosmic rays in the Universe.

 The author would like to thank H. Zaraket for critical questions,
 remarks and useful  references.

\newpage
\begin{figure} 
\begin{center}
\parbox[c]{13.5cm}{\epsfig{file=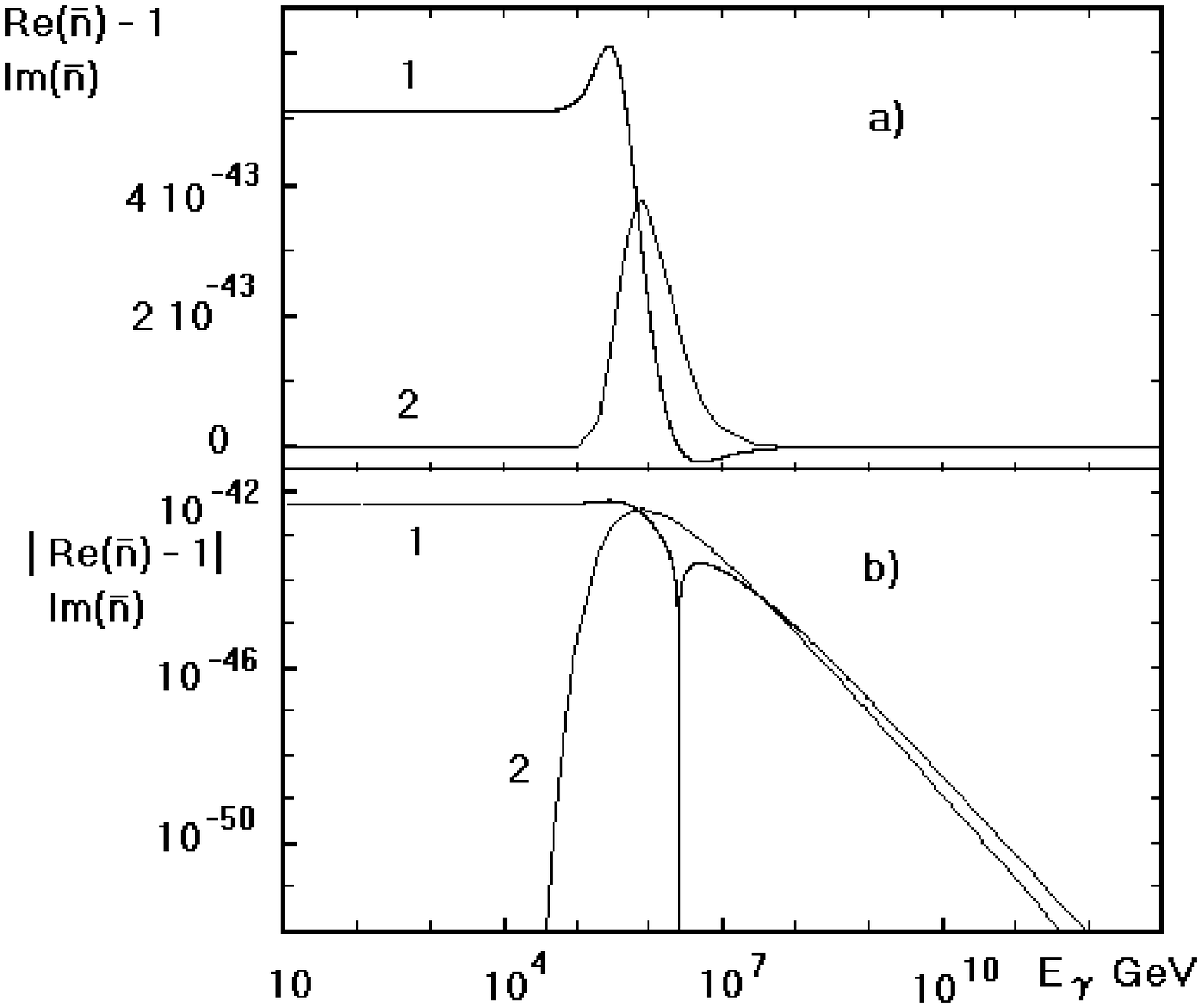,height=10.5cm}}
\parbox[c]{15cm}{\caption{
The real minus unit (1) and imaginary (2) parts of refractive index
(in linear (a) and logarithmic (b) scales) in
Cosmic Background Radiation  as functions of the photon energy.
 The absolute value $|Re{\tilde{n}}-1|$ is shown in logarithmic scale.
              }}  
\end{center} 
\end{figure}
\newpage
\begin{figure} 
\begin{center}
\parbox[c]{14.5cm}{\epsfig{file=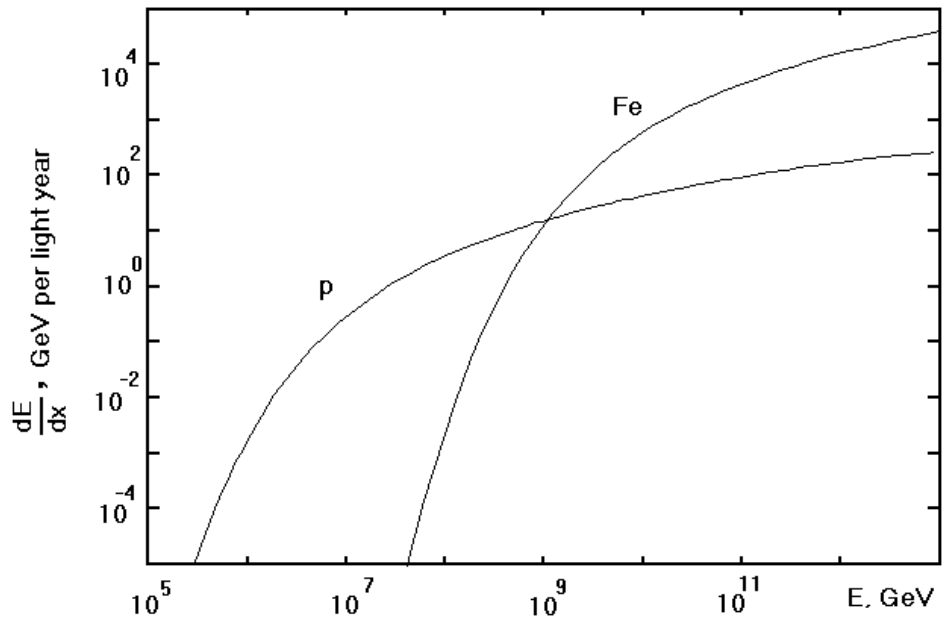,height=9cm}}
\parbox[c]{15cm}{\caption{
The vacuum polarization energy losses of  protons and iron nuclei
propagating in the extragalactic space as function of the energy.
              }}  
\end{center} 
\end{figure}
\begin{figure}
\begin{center}
\parbox[c]{13.5cm}{\epsfig{file=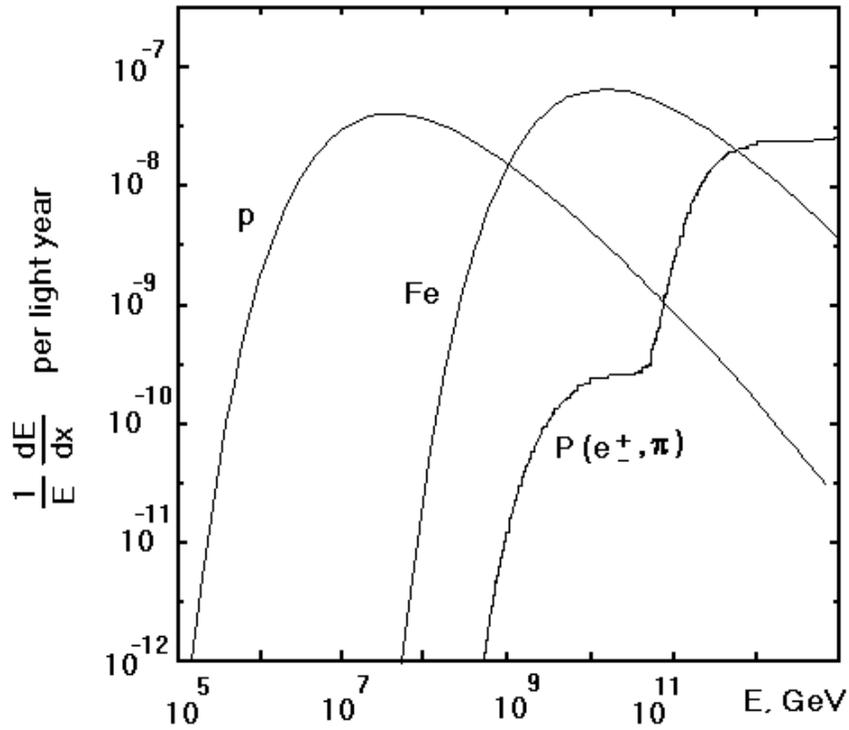,height=11cm}}
\parbox[c]{15cm}{\caption{
The relative vacuum polarization energy losses of  protons and iron nuclei
propagating in the extragalactic space as function of the energy.
On the right the similar value for photoproduction energy losses 
(from [2])
is presented  for comparison. 
}}  
\end{center} 
\end{figure}

\end{document}